\documentclass[conference]{IEEEtran}
\IEEEoverridecommandlockouts
\usepackage{cite}
\usepackage{amsmath,amssymb,amsfonts}
\usepackage{algorithmic}
\usepackage{graphicx}
\usepackage{textcomp}
\usepackage{xcolor}
\usepackage[nolist]{acronym}
\usepackage{paralist}
\usepackage{algorithm2e}
\def\BibTeX{{\rm B\kern-.05em{\sc i\kern-.025em b}\kern-.08em
    T\kern-.1667em\lower.7ex\hbox{E}\kern-.125emX}}

\begin{document}

\makeatletter

\algocf@newcmdside@kobe{Try@try}{%
    \KwSty{try}%
    \ifArgumentEmpty{#1}\relax{ #1}%
    \algocf@group{#2}%
    \par
}
\algocf@newcmdside@kobe{Try@except}{%
    \KwSty{except}%
    \ifArgumentEmpty{#1}\relax{ #1}%
    \algocf@block{#2}{end}{#3}%
    \par
}
\newcommand\Try[2]{%
    \Try@try{#1}%
    \Try@except{#2}%
}

\algocf@newcmdside@kobe{With@with}{%
    \KwSty{with}%
    \ifArgumentEmpty{#1}\relax{ #1}%
    \algocf@group{#2}%
    \par
}
\algocf@newcmdside@kobe{With@do}{%
    \KwSty{do}%
    \ifArgumentEmpty{#1}\relax{ #1}%
    \algocf@block{#2}{end}{#3}%
    \par
}
\newcommand\With[2]{%
    \With@with{#1}%
    \With@do{#2}%
}
\makeatother

\RestyleAlgo{ruled}

\begin{acronym}
\acro{sg}[SG]{smart grid}
\acroplural{sg}[SGs]{smart grids}
\acro{der}[DER]{distributed energy resource}
\acroplural{der}[DERs]{distributed energy resources}
\acro{ict}[ICT]{information and communication technology}
\acro{fdi}[FDI]{false data injection}
\acro{scada}[SCADA]{supervisory control and data acquisition}
\acro{dmz}[DMZ]{Demilitarized Zone}
\acro{mtu}[MTU]{master terminal unit}
\acroplural{mtu}[MTUs]{master terminal units}
\acro{hmi}[HMI]{human machine interface}
\acro{plc}[PLC]{programmable logic controller}
\acroplural{plc}[PLCs]{programmable logic controllers}
\acro{ied}[IED]{intelligent electronic device}
\acroplural{ied}[IEDs]{intelligent electronic devices}
\acro{rtu}[RTU]{remote terminal unit}
\acroplural{rtu}[RTUs]{remote terminal units}
\acro{iec104}[IEC-104]{IEC 60870-5-104}
\acro{apdu}[APDU]{application protocol data unit}
\acro{apci}[APCI]{application protocol control information}
\acro{asdu}[ASDU]{application service data unit}
\acro{io}[IO]{information object}
\acroplural{io}[IOs]{information objects}
\acro{cot}[COT]{cause of transmission}
\acro{mitm}[MITM]{Man-in-the-Middle}
\acro{fdi}[FDI]{false data injection}
\acro{ids}[IDS]{intrusion detection system}
\acroplural{ids}[IDSs]{intrusion detection systems}
\acro{siem}[SIEM]{security information and event management}
\acro{mv}[MV]{medium voltage}
\acro{lv}[LV]{low voltage}
\acro{pid}[pid]{process ID}
\acro{cdss}[CDSS]{controllable distribution secondary substation}
\acro{bss}[BSS]{battery storage system}
\acroplural{bss}[BSSs]{battery storage systems}
\acro{pv}[PV]{photovoltaic inverter}
\acro{mp}[MP]{measuring point}
\acroplural{mp}[MPs]{measuring points}
\acro{dsc}[DSC]{dummy SCADA client}
\acro{fcli}[FCLI]{Fronius CL inverter}
\acro{fipi}[FIPI]{Fronius IG+ inverter}
\acro{sii}[SII]{Sunny Island inverter}
\acro{tls}[TLS]{transport layer security}
\acro{actcon}[ActCon]{activation confirmation}
\acro{actterm}[ActTerm]{activation termination}
\acro{rtt}[RTT]{round trip time}
\acro{c2}[C2]{command and control}
\acro{dst}[DST]{Dempster Shafer Theory}
\acro{fcm}[FCM]{fuzzy cognitive map}
\acroplural{fcm}[FCMs]{fuzzy cognitive maps}
\acro{ec}[EC]{event correlator}
\acro{sc}[SC]{strategy correlator}
\acro{ioc}[IoC]{indicator of compromise}
\acro{hids}[HIDS]{host-based intrusion detection system}
\acro{nids}[NIDS]{network-based intrusion detection system}
\acroplural{ioc}[IoCs]{indicators of compromise}
\acro{ot}[OT]{operational technology}
\acro{it}[IT]{information technology}
\acro{iot}[IoT]{Internet-of-Things}
\acro{cvss}[CVSS]{common vulnerability scoring system}
\acro{cve}[CVE]{common vulnerability enumeration}
\acro{ip}[IP]{Internet protocol}
\acro{ems}[EMS]{energy management system}
\acro{hcpn}[HCPN]{hidden-colored-petri net}
\acro{dsr}[DSR]{demand side response}
\acro{cps}[CPS]{cyber-physical system}
\acroplural{cps}[CPSs]{cyber-physical systems}
\acro{pol}[PoL]{Pattern-of-Life}
\acro{ttp}[TTP]{tactic, technique and procedure}
\acroplural{ttp}[TTPs]{tactics, techniques and procedures}
\acro{oscti}[OSCTI]{open-source Cyber Threat Intelligence}
\acro{apt}[APT]{advanced persistent threat}
\end{acronym}

\bstctlcite{IEEEexample:BSTcontrol}

\title{Enhancing SCADA Security: Developing a Host-Based Intrusion Detection System to Safeguard Against Cyberattacks}

\author{
\IEEEauthorblockN{%
Ömer Sen\IEEEauthorrefmark{1}\IEEEauthorrefmark{2},
Tarek Hassan\IEEEauthorrefmark{1},
Andreas Ulbig\IEEEauthorrefmark{1}\IEEEauthorrefmark{2},
Martin Henze\IEEEauthorrefmark{3}\IEEEauthorrefmark{4},
}

\IEEEauthorblockA{%
\IEEEauthorrefmark{1}\textit{IAEW, RWTH Aachen University,} Aachen, Germany |
\IEEEauthorrefmark{2}\textit{Digital Energy, Fraunhofer FIT,} Aachen, Germany\\
Email: \{o.sen, a.ulbig\}@iaew.rwth-aachen.de, tarek.hassan1@rwth-aachen.de, \{oemer.sen, andreas.ulbig\}@fit.fraunhofer.de}
\IEEEauthorblockA{%
\IEEEauthorrefmark{3}\textit{SPICe, RWTH Aachen University,} Aachen, Germany |
\IEEEauthorrefmark{4}\textit{CA\&D, Fraunhofer FKIE,} Wachtberg, Germany\\
Email: henze@cs.rwth-aachen.de, martin.henze@fkie.fraunhofer.de}
}

\IEEEoverridecommandlockouts


\maketitle

\IEEEpubidadjcol

\begin{abstract}
With the increasing reliance of smart grids on correctly functioning SCADA systems and their vulnerability to cyberattacks, there is a pressing need for effective security measures.
SCADA systems are prone to cyberattacks, posing risks to critical infrastructure.
As there is a lack of host-based intrusion detection systems specifically designed for the stable nature of SCADA systems, the objective of this work is to propose a host-based intrusion detection system tailored for SCADA systems in smart grids.
The proposed system utilizes USB device identification, flagging, and process memory scanning to monitor and detect anomalies in SCADA systems, providing enhanced security measures. 
Evaluation in three different scenarios demonstrates the tool's effectiveness in detecting and disabling malware.
The proposed approach effectively identifies potential threats and enhances the security of SCADA systems in smart grids, providing a promising solution to protect against cyberattacks.

\end{abstract}

\begin{IEEEkeywords}
SCADA System, Cyber Security, Host-based Intrusion Detection System, Smart Grid, USB Device Malwalre
\end{IEEEkeywords}

\section{Introduction} \label{sec:introduction}
The integration of \ac{iot} with \ac{cps} has advanced operational capabilities in \ac{sg}, but it also introduces vulnerabilities as \ac{ot} and IT boundaries blur~\cite{babayigit2023industrial}.  However, the transition towards Industry 4.0 has encouraged the integration of cloud services and increased connectivity~\cite{dehlaghi2023icssim}. While this offers numerous advantages, it also exposes \ac{scada} systems to heightened cyber security risks.

The new threat landscape necessitate robust protective measures to safeguard against potential cyberattacks~\cite{khan2022enhancing}. The incorporation of \acp{dmz} in modern \ac{scada} architecture serves as a critical intermediary layer, mitigating risks from external networks as part of the defense-in-depth concept~\cite{wang2017security}. Despite existing security measures, these systems remain susceptible to indirect malware propagation. Challenges persist, especially concerning the insecurity of peripheral communication standards like USB and the potential risk of process memory hijacking~\cite{abou2021securing}. The historical instance of the Stuxnet attack, which utilized process hijacking and USB drive insertion as an attack vector~\cite{alanazi2023scada}, exemplifies this vulnerability and underscores the need for heightened security in \ac{scada} systems~\cite{van2020methods}. However, the focus of existing research has predominantly been on network-based \acp{ids}, often overlooking the critical need for \ac{scada}-specific \acp{hids} that are essential for addressing sophisticated, zero-day threats~\cite{alanazi2022scada}. However, these studies have not adequately addressed the use of USB device identification or process memory scanning and hashing in \ac{hids} for \ac{scada} systems.

Addressing the gaps in current methodologies, a comprehensive approach to enhance the security of \ac{scada} systems is needed. Such measures are critical in addressing the vulnerabilities posed by the process memory hijacking via insecure design of peripheral communication standards, thereby enhancing the overall security and resilience of \ac{scada} systems against a myriad of cyber threats~\cite{rrushi2022physics}. This oversight in the existing literature and the demonstrated vulnerabilities, such as those exploited by Stuxnet, highlight the critical need for a more comprehensive approach in \ac{hids} design for \ac{scada} systems.

In this paper we present our \ac{hids} approach, designed for the stable environments of \ac{scada} systems, that leverages their predictable behavior to enhance anomaly detection by targeting specific vulnerabilities like unauthorized hardware insertions. By integrating USB device identification, process memory scanning, and hashing, it ensures real-time, comprehensive protection against diverse cyber threats. Additionally, our specification-based approach for process monitoring, involving the definition of normal operational profiles and real-time monitoring, has been validated through case studies, demonstrating its effectiveness in practical \ac{scada} scenarios.

The key contributions of this work are:
\begin{enumerate}
    \item Providing a comprehensive overview of the current state of \ac{hids} in \ac{scada} systems, pinpointing the research gaps.
    \item Proposing a novel \ac{hids} framework tailored for \ac{scada} systems, focusing on protecting against unknown threats via hardware and memory-based attacks.
    \item Validating the proposed \ac{hids} approach through various scenarios, demonstrating its effectiveness and reliability in real-world applications.
\end{enumerate}

This paper is structured as follows: Section 1 introduces \ac{scada} systems' vulnerabilities; 
Section 2 discusses the proposed \ac{hids} design; 
Section 3 details the case studies and discussion; 
followed by concluding remarks in Section 5.

\section{Multi-Staged Attack Detection System} \label{sec:framework}
In this section, we present our proposed \ac{hids} approach and the method we use in malware detection.
\subsection{Overview of HIDS Approach}
\begin{figure*}
    \centerline{\includegraphics[width=2\columnwidth, height=0.2\linewidth]{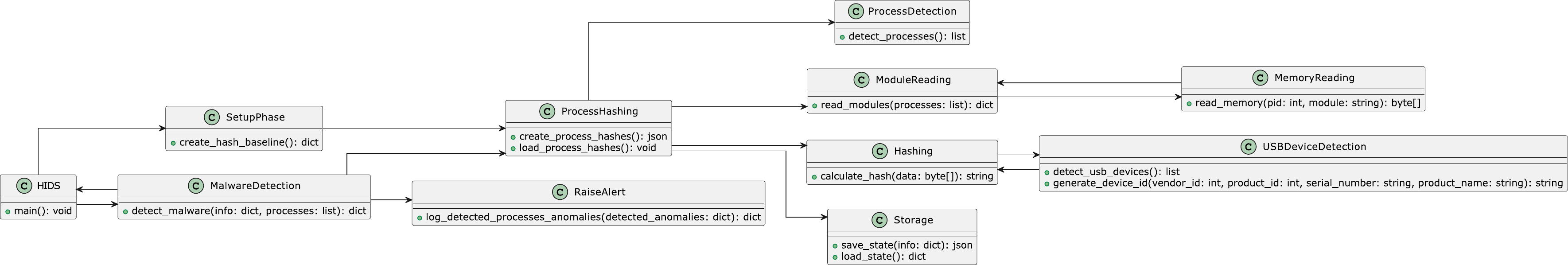}}
    \caption{Component diagram depicting the proposed multi-stage \ac{hids} for \ac{scada} systems, which combines USB device identification, process memory scanning, and hashing techniques to detect and disable malware attacks, ensuring real-time protection and comprehensive security coverage. 
    }
    \label{fig:framework_overview}
    \vspace{-1em}
\end{figure*}

The proposed method outlines the development of a \ac{hids} specifically designed for \ac{scada} systems (cf. Figure~\ref{fig:framework_overview}).
The \ac{hids} combines USB device identification, process memory scanning, and hashing techniques to detect and disable malware attacks on \ac{scada} systems.
It follows a multi-stage approach, involving multiple components working together to provide comprehensive security coverage.
By integrating these components, the \ac{hids} aims to detect anomalies and potential security threats in real-time, ensuring the robust protection of \ac{scada} systems.
By compiling the Python scripts with the runtime interpreter for the dedicated operating system, a portable executable \ac{hids} for Linux-based \ac{scada} systems can be created.

\subsection{Setup Phase for Data Collection}
In this section, we describe the setup phase for the data collection process of the proposed \ac{hids} approach for later use in malware detection.

In the setup phase of our \ac{hids}, tailored for \ac{scada} systems, we create a secure baseline encompassing both a comprehensive whitelist of USB devices and a catalog of legitimate processes and modules. 
This involves meticulously vetting USB devices and running processes with unique identifiers for legitimate devices and process memory hashes, while cryptographic methods and strict access controls safeguard this baseline, ensuring protection against unauthorized modifications during operation.

\subsubsection{USB Device Identification}
The \ac{hids} incorporates USB device identification as a critical component of its approach to enhance security.
USB devices have the potential to serve as entry points for malicious actors, making their detection and monitoring crucial for system protection.
The approach utilizes Python-based libraries, specifically $usb.core$ and $usb.util$, to retrieve essential information about connected USB devices.
These libraries facilitate the generation of unique device IDs, which are then compared against a trusted ID list to identify potentially harmful devices.
Unauthorized devices are flagged and disabled by detaching them from the kernel driver.
By leveraging the capabilities of the $usb.core$ and $usb.util$ libraries, the approach offers an effective and comprehensive solution for USB device detection and security monitoring (cf. Algorithm~\ref{alg:usb-identification}).
\begin{algorithm}
\caption{USB Device Identification}\label{alg:usb-identification}
\SetKwProg{Fn}{Function}{:}{}
\SetKwInOut{Input}{Input}
\SetKwInOut{Output}{Output}
\footnotesize

\Fn{check\_usb()}{
  devices $\gets$ \textbf{find\_all\_usb\_devices}()\;

  \For{device \textbf{in} devices}{
    vendor\_id $\gets$ \textbf{get\_vendor\_id}(device)\;
    product\_id $\gets$ \textbf{get\_product\_id}(device)\;
    product\_name $\gets$ \textbf{get\_product\_name}(device)\;
    serial\_number $\gets$ \textbf{get\_serial\_number}(device)\;

    device\_id $\gets$ generate\_id(vendor\_id, product\_id, serial\_number, product\_name)\;

    \If{device\_not\_in\_allow\_list(device\_id)}{
      \textbf{detach\_kernel\_driver}(device)\;
    }
  }
}

\end{algorithm}
The USB Device Identification algorithm is designed to identify and manage connected USB devices by generating unique device IDs and verifying them against an allow list.
The algorithm begins by searching for all connected USB devices and extracting essential information such as the vendor ID, product ID, product name, and serial number. 
This information is used to create a unique device identifier.

Next, the algorithm checks if the generated device ID is present in the allow list. 
If the device is not found in the allow list, it is considered potentially unauthorized or malicious. 
In such cases, the algorithm takes proactive measures by detaching the device from the kernel driver, effectively disabling it.

By applying these steps to each connected USB device, the algorithm ensures that only authorized devices remain active. 
It provides a mechanism to identify and take action against potentially harmful or unauthorized USB devices within a system. 
This approach extends the operational capabilities of a conventional \ac{hids}, which typically operates passively and only raises alerts.

\subsubsection{Process Memory Scanning and Hashing}
The \ac{hids} employs process memory scanning and hashing techniques to detect malicious activity within running processes.
The approach involves reading and loading all processes on the system, using the $psutil$ library in Python.
Malware can infect processes through various techniques, posing a significant threat to \ac{scada} systems.
The process memory scanning and hashing feature focuses on the executable code section of processes, which contains critical instructions for program behavior.
By comparing generated hashes of process memory, the \ac{hids} can flag abnormal memory regions for further analysis.
The implementation includes reading process memory using specific files in Linux systems and utilizing the SHA-256 hashing algorithm to generate unique hashes (cf. Algorithm~\ref{alg:hids-algo-1}).
\begin{algorithm}
\caption{\ac{hids} Algorithm Process Hashing}\label{alg:hids-algo-1}
\SetKwProg{Function}{Function}{:}{}
\SetKwInOut{Input}{Input}
\SetKwInOut{Output}{Output}
\footnotesize

\Function{get\_module\_hash(module\_dump)}{
    \Input{module\_dump}
    \Output{sha256\_result}

    m $\gets$ \textbf{hash\_sha256}()\;
    \textbf{update} m \textbf{to} module\_dump\;
    sha256\_result $\gets$ \textbf{Hash\_value\_repres}(m)\;
    \textbf{return} sha256\_result\;
}

\Function{read\_module\_memory(pid, module\_path)}{
    \Input{pid, module\_path}
    \Output{module\_dump}

    module\_dump $\gets$ byte\_array()\;
    maps\_file $\gets$ \textbf{read}('/proc/{pid}/maps')\;
    mem\_file $\gets$ \textbf{read}('/proc/{pid}/mem')\;

    \ForEach{line \textbf{in} maps\_file}{
    
        \If{module\_path \textbf{not in} line}{
            \textbf{continue with next iteration}\;
        }

        m $\gets$ \textbf{get\_memory\_region\_address}(line)\;
        s\_addr $\gets$ \textbf{start\_address}(m)\;
        e\_addr $\gets$ \textbf{end\_address}(m)\;
        perm $\gets$ \textbf{permissions}(m)\;

        \If{invalid\_memory\_region(s\_addr) \textbf{or} not\_executable(perm)}{
            \textbf{continue with next iteration}\;
        }

        \textbf{seek} s\_addr \textbf{in} mem\_file\;
        size $\gets$ e\_addr - s\_addr\;
        chunk $\gets$ \textbf{read\_size}(mem\_file)\;
        \textbf{extend} module\_dump \textbf{with} chunk\;
    }

    \Return module\_dump\;
}
\end{algorithm}
The process detection and hashing algorithm checks the memory of running processes to gather information and calculate hashes. 
It utilizes the $read\_module\_memory$ function, which reads the memory regions of a process specified by its \ac{pid} and module path. 
The function extracts relevant information, such as start and end addresses and permissions, from the process's memory map file. 
It then checks if the region is readable and executable before reading the actual memory from the process's memory file. 
The collected memory is stored in a bytearray called $module_dump$. The $get\_module\_hash$ function uses the $hashlib.sha256$ function to calculate the SHA-256 hash of the module's memory dump. 
It returns the hexadecimal representation of the hash.

This algorithm allows for the identification of changes or anomalies in the memory of running processes by examining their memory maps and extracting the executable memory regions. 
The calculated hashes of the module's memory facilitate the comparison and detection of potential malicious activity. Algorithm~\ref{alg:hids-algo-2} summarizes the entire setup phase of the data collection process.
%
\begin{algorithm}
\caption{\ac{hids} Algorithm Setup Phase}\label{alg:hids-algo-2}
\SetKwProg{Function}{Function}{:}{}
\SetKwInOut{Input}{Input}
\SetKwInOut{Output}{Output}
\footnotesize

\Function{start\_setup\_phase()}{
    info $\gets$ key\_value\_map()\;
    info['module\_hashes'] $\gets$ key\_value\_map()\;
    info['process\_modules'] $\gets$ key\_value\_map()\;

    \ForEach{proc \textbf{in} \textbf{iterate\_processes}()}{
        proc\_name $\gets$ \textbf{get\_name}(proc)\;
    
        info['process\_modules'][proc\_name] $\gets$ list()\;
        info['module\_hashes'][proc\_name] $\gets$ key\_value\_map()\;

        \If{proc\_name \textbf{not in} info['process\_modules']}{
            \textbf{log}('New process found:' proc\_name)\;
            \textbf{continue with next iteration}\;
        }

        modules $\gets$ memory\_maps(proc)\;

        \ForEach{module \textbf{in} modules}{
            module\_path $\gets$ \textbf{get\_path}(module)
        
            \If{special\_process\_regions(module)}{
                \textbf{continue with next iteration}\;
            }

            \textbf{append} info['process\_modules'][proc\_name] \textbf{with} module\_path\;
            module\_dump $\gets$ read\_module\_memory(proc\_name, module\_path)\;

            \If{module\_path \textbf{not in} info['module\_hashes'][proc\_name]}{
                info['module\_hashes'][proc\_name] [module\_path] $\gets$ get\_module\_hash(module\_dump)\;
            }
        }
        \textbf{dump} info \textbf{in} 'data\_storage'\;
    }
}
\end{algorithm}
\subsection{Malware Detection}
The \ac{hids} approach aims to provide effective malware detection (cf. Algorithm~\ref{alg:hids-algo-3}) within running processes using a baseline created during the setup phase. 
By combining USB device identification and process memory scanning, the \ac{hids} can detect potential malware threats and abnormal activities.
The \ac{hids} architecture includes components for tracking and monitoring running processes, checking loaded modules, and comparing hashes.
We focus on identifying unauthorized changes and anomalies in the executable memory sections of processes, providing comprehensive security coverage against various malware infection techniques.
\begin{algorithm}
\caption{\ac{hids} Algorithm Malware Detection}\label{alg:hids-algo-3}
\SetKwProg{Function}{Function}{:}{}
\SetKwInOut{Input}{Input}
\SetKwInOut{Output}{Output}
\footnotesize

\Function{check\_processes(info)}{
    \Input{info}

    module\_hashes $\gets$ info['module\_hashes']\;

    \ForEach{proc \textbf{in} \textbf{iterate\_processes}()}{
        proc\_name $\gets$ \textbf{get\_name}(proc)\;
        pid $\gets$ \textbf{get\_pid}(proc)
    
        modules $\gets$ \textbf{memory\_maps(proc)}\;

        \ForEach{module \textbf{in} modules}{
            module\_path $\gets$ \textbf{get\_path}(module)
            
            \If{special\_process\_regions(module)}{
                \textbf{continue with next iteration}\;
            }

            \If{module\_path \textbf{not in} info['process\_modules'][proc\_name]}{
                \textbf{log}('UNKNOWN MODULE FOUND'  module\_path 'in' proc\_name)\;
            }

            \eIf{module.path \textbf{in} module\_hashes[proc.\textbf{name()}]}{
                module\_dump $\gets$ read\_module\_memory(pid, module\_path)\;
                sha256\_result $\gets$ get\_module\_hash(module\_dump)\;

                \If{module\_hashes[proc\_name][module\_path] \textbf{$\neq$} sha256\_result}{
                    \textbf{log}('FAILED' module\_path 'in' proc\_name)\;
                }
            }{
                \textbf{log}('UNKNOWN MODULE FOUND' module\_path 'in' proc\_name)\;
            }
        }
    }
}
\end{algorithm}
The proposed algorithm employs a process hashing technique to ensure the integrity of running processes in the system. 
It consists of two key functions: $get\_module\_hash$ and $read\_module\_memory$. The former calculates a unique identifier for each module by computing its SHA256 hash based on its memory dump. 
The latter reads and filters the memory regions of a specific process, extracting the executable regions and assembling them into a byte array, representing the module dump.

During the setup phase, the algorithm iterates over all running processes, collecting information about their modules and their respective hashes. 
If a new process is detected, it is added to the information dictionary. 
Non-executable or irrelevant regions are filtered out, and the $read\_module\_memory$ function is used to obtain the module dump. 
The calculated module hash is then stored in the information dictionary. 
The collected data is saved for later use.

The Process Checking phase periodically examines the running processes to detect any changes or discrepancies in their module hashes. 
The stored information is loaded, and the algorithm compares the stored hashes with the recalculated hashes of the modules. 
If a module is not found in the stored information, it is flagged as an ''UNKNOWN MODULE''. 
If the stored and recalculated hashes differ, it suggests a potential intrusion or modification, and the module is flagged as ''FAILED''. 
The results of these checks are displayed for monitoring purposes.

The main function serves as the entry point of the algorithm, initiating an infinite loop that continuously performs process checking at regular intervals. 
This ensures real-time monitoring and detection of any suspicious activities.

In summary, the \ac{hids} algorithm utilizes process hashing, setup phase, and process checking to provide a mechanism for identifying unauthorized or malicious modifications in the memory regions of running processes, thereby ensuring the security and integrity of the system.


\section{Case Study \& Discussion} \label{sec:result}
To investigate the detection quality of our proposed framework, we evaluate it using a case study.
\subsection{Investigation Procedure Setup}
Our experimental validation of the \ac{hids} for \ac{scada} systems included scenarios that closely mirrored real-world conditions, focusing on the system's ability to detect and disable malware introduced through USB devices and other attack vectors. 
The investigation environment comprised the \ac{scada} system, the \ac{hids}, two processes (one legitimate and one malicious), and an \ac{rtu}, running on a Linux host within a \ac{dmz}. 
We specifically chose unauthorized devices and programs for our testing based on their known susceptibility to USB-based malware, a critical vulnerability in \ac{scada}  systems. 
The experiment involved simulating attacks using USB devices equipped with autorun scripts, emulating common malware tactics, to rigorously test the \ac{hids}'s detection and response capabilities. 
In these scenarios, the \ac{hids} was tasked with monitoring activities such as USB device insertion and program execution, employing USB device identification, process memory scanning, and hashing techniques to effectively identify and counteract malware threats. 
This approach provided a realistic and controlled setting to evaluate the \ac{hids}'s performance, ensuring a comprehensive assessment of its capability to defend against diverse cyber attacks in \ac{scada}  environments.

The evaluation process involves three specific scenarios. 
\begin{figure}
    \centerline{\includegraphics[width=\columnwidth]{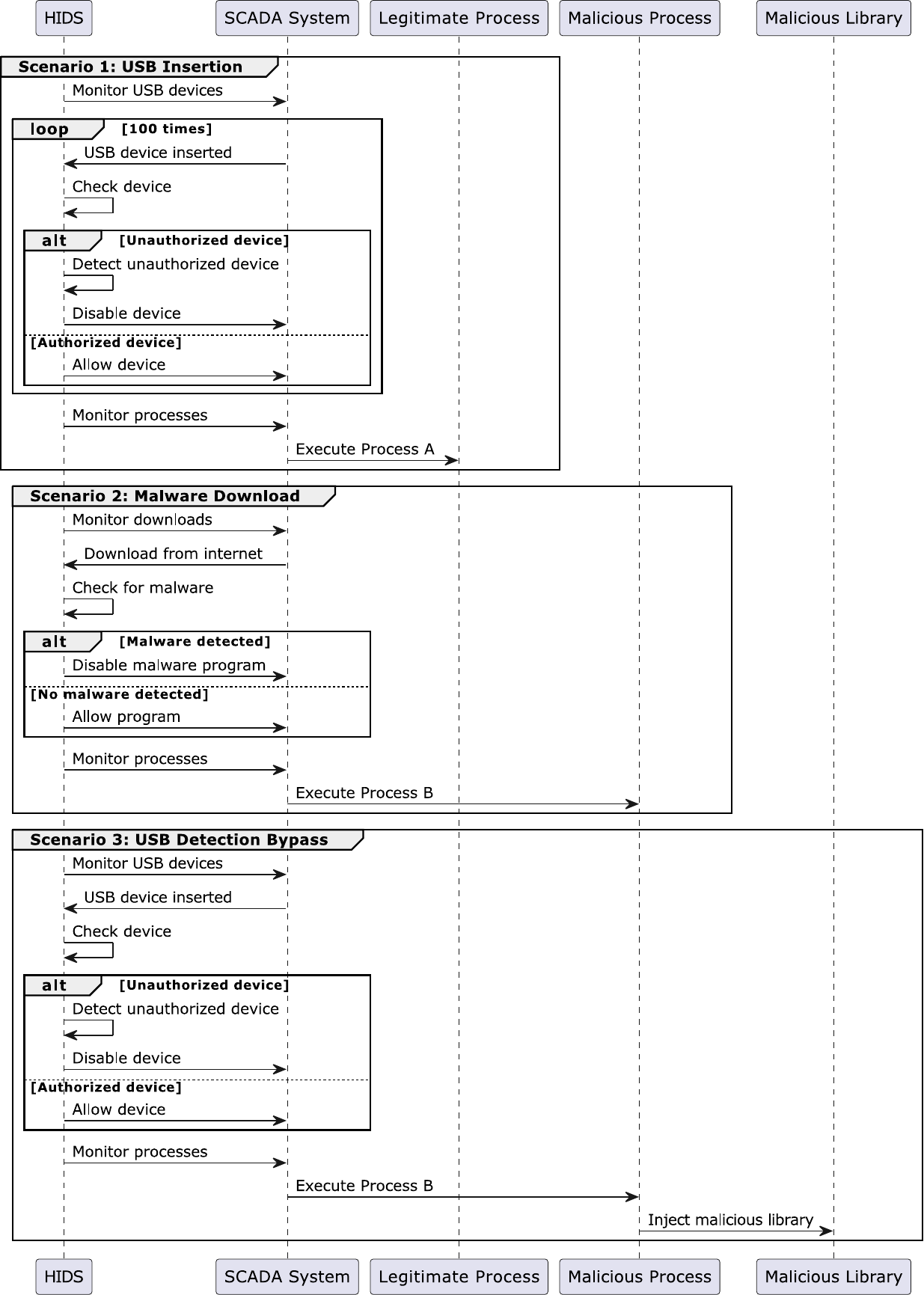}}
    \caption{Investigation environment depicting the assessment of \ac{scada} system vulnerability and the proposed \ac{hids} through three scenarios: USB device insertion, downloaded malware, and USB detection bypass. 
    }
    \label{fig:res_investigation_env}
    \vspace{-1em}
\end{figure}
In our comprehensive experimental validation of the \ac{hids} for \ac{scada} systems, we conducted three distinct scenarios to assess its efficacy against various malware threats. 
Scenario 1 focused on unauthorized USB devices, where we tested the \ac{hids}'s speed and accuracy in detecting and disabling malware, simulating this by inserting a USB device into the \ac{scada} computer 100 times. 
Scenario 2 and 3 further expanded our evaluation, with the former assessing the \ac{hids}'s response to malware downloaded from the Internet and the latter examining its ability to detect malware from undetected USB devices. 
Each of these scenarios, repeated 100 times, provided a controlled yet realistic environment to rigorously test the \ac{hids}'s capabilities in identifying and neutralizing a wide range of malware attacks in \ac{scada} environments, reflecting actual operational challenges.
\subsection{Results}
The results of each scenario are presented below:
\paragraph{Scenario 1: USB Device Insertion}
In the first scenario, the proposed \ac{hids} reached a 100\% detection rate for unauthorized USB devices. 
However, the disabling rate was found to be 95\%, indicating that in some instances, the tool was unable to disable the devices before they could cause harm.  
Further analysis identified a race condition between the autorun function and the \ac{hids}'s detection mechanism as the primary cause of this limitation. 
\paragraph{Scenario 2: Downloaded Malware}
The \ac{hids} demonstrated remarkable effectiveness in detecting malware introduced through downloaded programs from the Internet. 
It achieved a 100\% detection rate by accurately identifying changes in process memory caused by the malware program. 
This highlights the tool's capability to counter sophisticated malware attacks that often employ techniques to evade traditional antivirus software.
\paragraph{Scenario 3: USB Detection Bypass}
In the third scenario, the \ac{hids} proved highly effective in detecting malware introduced through USB devices. 
It successfully identified foreign modules loaded into the processes of the \ac{scada} system, even when the malware bypassed the prevention phase of USB device disabling. 
This showcases the tool's potential to defend against diverse malware attacks and enhance the security of \ac{scada} systems. Thus, the detection rate was also 100\%. 
\subsection{Discussion}
The investigation revealed the strengths and limitations of the proposed \ac{hids} in detecting and disabling malware in \ac{scada} systems. 
While the tool demonstrated a high detection rate for unauthorized USB devices and malware introduced through downloads and undetected USB devices, there were areas for improvement.
Another observation is the rate with which unauthorized USB devices are immediately disabled in scenario 1. 
The race condition between the autorun function and the \ac{hids}'s detection mechanism resulted in a 95\% disabling rate. 
To address this, future research should focus on refining the detection mechanism and minimizing the impact of race conditions, thereby enhancing the tool's ability to disable unauthorized devices efficiently.
Another challenge is the automatic program updates that lead to unnecessary alerts due to changes in the executable memory sections. 
Additionally, relocatable code within some processes can contribute to the false detection of executable memory modification. 
To mitigate these issues, proposed solutions include creating a whitelist of known updates, implementing a versioning system, distinguishing between relocatable code and modified code, and monitoring process behavior for suspicious activity.
Overall, the investigation highlighted the effectiveness of the proposed \ac{hids} in detecting and mitigating malware in \ac{scada} systems. 
The results emphasize the need for ongoing research to address the identified limitations and further refine the tool's capabilities. 
It is crucial to develop alternative methods and strategies to enhance the \ac{hids}'s performance and strengthen the defense against evolving malware attacks in \ac{scada} systems.

\section{Conclusion} \label{sec:conclusion}
The increasing security concerns surrounding \ac{scada} systems in \ac{sg}, which control critical industrial processes, necessitate the development of effective security solutions. 
The objective of this work was to propose a novel \ac{hids} capable of detecting and disabling malware attacks on \ac{scada} systems in \ac{sg}.
The proposed \ac{hids} utilized a combination of USB device security features and process memory scanning algorithms to monitor and analyze specific aspects of the system, providing comprehensive security coverage. 
The investigation revealed the strengths and limitations of the proposed \ac{hids} in detecting and disabling malware in \ac{scada} systems. While effective in detecting malware, challenges such as the race condition in scenario 1 and false positives due to automatic updates were identified. 
%
%
Future research should aim to refine the approach's detection mechanisms through advanced algorithms for real-time malware identification. 
Additionally, exploring the integration of machine learning and hardware-based intrusion detection systems is recommended for a more comprehensive defense against diverse malware attacks.


\bibliographystyle{IEEEtran}
\bibliography{conference_101719}

\begin{thebibliography}{1}
\providecommand{\url}[1]{#1}
\csname url@samestyle\endcsname
\providecommand{\newblock}{\relax}
\providecommand{\bibinfo}[2]{#2}
\providecommand{\BIBentrySTDinterwordspacing}{\spaceskip=0pt\relax}
\providecommand{\BIBentryALTinterwordstretchfactor}{4}
\providecommand{\BIBentryALTinterwordspacing}{\spaceskip=\fontdimen2\font plus
\BIBentryALTinterwordstretchfactor\fontdimen3\font minus
  \fontdimen4\font\relax}
\providecommand{\BIBforeignlanguage}[2]{{%
\expandafter\ifx\csname l@#1\endcsname\relax
\typeout{** WARNING: IEEEtran.bst: No hyphenation pattern has been}%
\typeout{** loaded for the language `#1'. Using the pattern for}%
\typeout{** the default language instead.}%
\else
\language=\csname l@#1\endcsname
\fi
#2}}
\providecommand{\BIBdecl}{\relax}
\BIBdecl

\bibitem{babayigit2023industrial}
B.~Babayigit \emph{et~al.}, ``Industrial internet of things: A review of
  improvements over traditional scada systems for industrial automation,''
  \emph{IEEE Systems Journal}, 2023.

\bibitem{dehlaghi2023icssim}
A.~Dehlaghi-Ghadim \emph{et~al.}, ``Icssim—a framework for building
  industrial control systems security testbeds,'' \emph{Computers in Industry},
  vol. 148, p. 103906, 2023.

\bibitem{khan2022enhancing}
I.~A. Khan \emph{et~al.}, ``Enhancing iiot networks protection: A robust
  security model for attack detection in internet industrial control systems,''
  \emph{Ad Hoc Networks}, 2022.

\bibitem{wang2017security}
S.~P. Wang \emph{et~al.}, ``Security by design: Defense-in-depth iot
  architecture,'' in \emph{Journal of The Colloquium for Information Systems
  Security Education}, vol.~4, no.~2, 2017, pp. 15--15.

\bibitem{abou2021securing}
A.~Abou~el Kalam, ``Securing scada and critical industrial systems: From needs
  to security mechanisms,'' \emph{IJoCIP}, 2021.

\bibitem{alanazi2023scada}
M.~Alanazi \emph{et~al.}, ``Scada vulnerabilities and attacks: A review of the
  state-of-the-art and open issues,'' \emph{Computers \& Security}, vol. 125,
  p. 103028, 2023.

\bibitem{van2020methods}
D.~van~der Velde \emph{et~al.}, ``Methods for actors in the electric power
  system to prevent, detect and react to ict attacks and failures,'' in
  \emph{ENERGYCon}, 2020.

\bibitem{alanazi2022scada}
M.~Alanazi \emph{et~al.}, ``Scada vulnerabilities and attacks: A review of the
  state-of-the-art and open issues,'' \emph{Computers \& Security}, 2022.

\bibitem{rrushi2022physics}
J.~L. Rrushi, ``Physics-driven page fault handling for customized deception
  against cps malware,'' \emph{TECS}, 2022.

\end{thebibliography}

\end{document}